\documentclass[12pt]{article}
\usepackage{graphicx}

\begin{document}

\title{Electromagnetic form factors of nucleons and $p\rightarrow\Delta$
\footnote{\bf \large Published in Physica Sinica, Vol.24, No.2, 124, March 1975
(Chinese, translated by Jun Gao)}}

\author{Bing An Li \\
Institute of High Energy Physics, Academica Sinica \\ Beijing,
P.R. China
\footnote{permanent address: Dept. of Phys., Univ. of Kentucky, Lexington, KY 40506}\\\\\\\\\\\\}

\maketitle

\pagebreak
\begin{abstract}
A relativistic quark model and a new set of wave functions of nucleon and $\Delta$
have been
used to study the electromagnetic properties of $\frac 12^{+}$
baryons and photoelectric production of $\Delta (1236)$.
Theoretical results of
$G_M^p(q^2)$, $\mu _p$, $\mu _\Lambda $, $\mu
_pG_E^p(q^2)/G_M^p(q^2)$, $G_E^n(q^2)$, and
$G_{M1+}(q^2)$, $\mu $, $E_{1+}$, $S_{1+}$ of
$p\rightarrow \Delta$ are presented.
\end{abstract}

\pagebreak

\section{Introduction}
The relativistic quark model [1] is successful in studying the electromagnetic
and weak interactions of ground-state baryons and mesons.
However, some results are
inconsistent with experimental data[3], for instance, theoretical ratio
$\mu _pG_E^p(q^2)/G_M^p(q^2)$ drops faster with $q^2$ than
experimental values. On the other hand, recently there has been
some experimental results about the electromagnetic properties of
ground-state baryons, for example, the measurement [4,5] of the
magnetic transition form factor $G_{M1+}(q^2)$ for $p\rightarrow
\Delta ^{+}(1236)$. These results need theoretical explanation. In
Ref.[2], a new set of baryon wave
functions have been constructed by requiring $SU(6)$ symmetry in the frame of center of mess.
For example, in the wave function of $\frac 12^{+}$ baryon there are
additional terms
\begin{equation}
\{(1+\frac im\widehat{p})\gamma _5C\}_{\alpha \beta }u_\lambda (p)_\gamma
+2C_{\alpha \beta }\{\gamma _5u_\lambda (p)\}_\gamma .
\end{equation}
p is the momentum of the baryon, m is the mass,
and $C=i\gamma _2\gamma _4$ which is the charge-conjugation
operator. The new wavefunctions are still s-wave
and satisfy $SU(6)$ symmetry
in the frame of center of mass. It is interesting to point out that the original
wave function
\[\{(1-\frac im\widehat{p})\gamma _5C\}_{\alpha \beta }u_\lambda (p)_\gamma\]
is constructed by the spinors of quarks with zero momentum and the new terms(1)
are constructed by the spinors of antiquarks with zero momentum. Both satisfy
$SU(6)$ in the frame of center of mass.

In this paper, the method proposed in Ref.[1] and the wave
functions constructed in Ref. [2] are used to study
the electromagnetic properties of nucleons and $p\rightarrow \Delta
^{+}(1236)$.
\section{Matrix element of electric currents}

The effective Hamiltonian of
electromagnetic interaction in quark model [1,2] is
\begin{equation}
H_i(x)=-ie\overline{\psi }(x)Q\{\widehat{A}(x)-\frac{i\kappa }{4m_p}\sigma
_{\mu \nu }F_{\mu \nu }(x)\}\psi (x).
\end{equation}
The wave function of $\frac 12^{+}$ baryon is[2]
\begin{eqnarray}
B_{\alpha \beta \gamma ,ijk}^{\frac 12\lambda }(x_1,x_2,x_3)_l^{l^{^{\prime
}}} &=&\frac 1{6\sqrt{2}}\sqrt{\frac mE}\varepsilon _{i^{^{\prime
}}j^{^{\prime }}k^{^{\prime }}}\{\Gamma _{\alpha \beta ,\gamma }(p)_\lambda
(\varepsilon _{ijl^{^{\prime }}}\delta _{kl}+\varepsilon _{ikl^{^{\prime
}}}\delta _{jl})  \nonumber \\
&&+\Gamma _{\beta \gamma ,\alpha }(p)_\lambda (\varepsilon _{jkl^{^{\prime
}}}\delta _{il}+\varepsilon _{ikl^{^{\prime }}}\delta _{jl})\},  \nonumber \\
\overline{B}_{\alpha \beta \gamma ,ijk}^{\frac 12\lambda
}(x_1,x_2,x_3)_l^{l^{^{\prime }}} &=&-\frac 1{6\sqrt{2}}\sqrt{\frac mE}%
\varepsilon _{i^{^{\prime }}j^{^{\prime }}k^{^{\prime }}}\{\overline{\Gamma }%
_{\alpha \beta ,\gamma }(p)_\lambda (\varepsilon _{ijl^{^{\prime }}}\delta
_{kl}+\varepsilon _{ikl^{^{\prime }}}\delta _{jl})  \nonumber \\
&&+\overline{\Gamma }_{\beta \gamma ,\alpha }(p)_\lambda (\varepsilon
_{jkl^{^{\prime }}}\delta _{il}+\varepsilon _{ikl^{^{\prime }}}\delta
_{jl})\}.
\end{eqnarray}
\begin{eqnarray}
\Gamma _{\alpha \beta ,\gamma }(p)_\lambda  &=&\{[f_1(x_1,x_2,x_3)-\frac im%
\widehat{p}f_2(x_1,x_2,x_3)]\gamma _5C\}_{\alpha \beta }u_\lambda (p)_\gamma
\nonumber \\
&&+\{f_1(x_1,x_2,x_3)-f_2(x_1,x_2,x_3)\}C_{\alpha \beta }\{\gamma
_5u_\lambda (p)\}_\gamma ,  \nonumber \\
\overline{\Gamma }_{\alpha \beta ,\gamma }(p)_\lambda
&=&\{C[f_1(-x_1,-x_2,-x_3)+\frac im\widehat{p}f_2(-x_1,-x_2,-x_3)]\gamma
_5\}_{\alpha \beta }\overline{u}_\lambda (p)_\gamma   \nonumber \\
&&+\{f_1(-x_1,-x_2,-x_3)-f_2(-x_1,-x_2,-x_3)\}C_{\alpha \beta }\{\overline{u}%
_\lambda (p)\gamma _5\}_\gamma.
\end{eqnarray}
The wave function of $\frac 32^{+}$ baryon is
\begin{eqnarray}
B_{\alpha \beta \gamma ,ijk}^{\frac 32\lambda ,lmn}(x_1,x_2,x_3) &=&\frac 1{2%
\sqrt{2}}\sqrt{\frac mE}\varepsilon _{i^{^{\prime }}j^{^{\prime
}}k^{^{\prime }}}d_{ijk}^{lmn}\Gamma _{\alpha \beta \gamma }(p)_\lambda ,
\nonumber \\
\overline{B}_{\alpha \beta \gamma ,ijk}^{\frac 32\lambda ,lmn}(x_1,x_2,x_3)
&=&\frac 1{2\sqrt{2}}\sqrt{\frac mE}\varepsilon _{i^{^{\prime }}j^{^{\prime
}}k^{^{\prime }}}d_{ijk}^{lmn}\overline{\Gamma }_{\alpha \beta \gamma
}(p)_\lambda .
\end{eqnarray}
\begin{eqnarray}
\Gamma _{\alpha \beta \gamma }(p)_\lambda  &=&\{[f_2(x_1,x_2,x_3)-\frac im%
\widehat{p}f_1(x_1,x_2,x_3)]\gamma _\mu C\}_{\alpha \beta }\psi _\mu
^\lambda (p)_\gamma   \nonumber \\
&&+\frac im\{f_1(x_1,x_2,x_3)-f_2(x_1,x_2,x_3)\}\{\gamma _\mu \widehat{p}%
\gamma _5C\}_{\alpha \beta }\{\gamma _5\psi _\mu ^\lambda (p)\}_\gamma ,
\nonumber \\
\overline{\Gamma }_{\alpha \beta \gamma }(p)_\lambda
&=&\{C[f_2(-x_1,-x_2,-x_3)+\frac im\widehat{p}f_1(-x_1,-x_2,-x_3)]\gamma
_\mu \}_{\alpha \beta }\overline{\psi }_\mu ^\lambda (p)_\gamma   \nonumber
\\
&&+\frac im\{f_1(-x_1,-x_2,-x_3)-f_2(-x_1,-x_2,-x_3)\}  \nonumber \\
&&\times \{C\widehat{p}\gamma _\mu \gamma _5\}_{\alpha \beta }\{\overline{%
\psi }_\mu ^\lambda (p)\gamma _5\}_\gamma ,
\end{eqnarray}
where $i^{^{\prime }}j^{^{\prime }}k^{^{\prime }}$ are color indices,
$f_{1,2}(x_1,x_2,x_3)$ are two
Lorentz-invariant spacial functions, p is the momentum of baryon. The wave functions(3-6) satisfy $SU(6)$ symmetry in the frame of center of mass[2].
They are s-wave in the rest frame.

For the model with three degenerate
states [6], the electric charge operator can be written as
\begin{equation}
Q_{k_1k_2}^{k_1^{^{\prime }}k_2^{^{\prime }}}=\delta _{k_11}\delta
_{k_21}\delta _{k_1^{^{\prime }}k_2^{^{\prime }}}-\delta
_{k_1k_2}\delta _{k_1^{^{\prime }}3}\delta _{k_2^{^{\prime }}3}
\end{equation}
and the following relationship is obtained
\begin{equation}
\frac 16\varepsilon _{k_1^{^{\prime }}j^{^{\prime }}i^{^{\prime
}}}\varepsilon _{k_2^{^{\prime }}j^{^{\prime }}i^{^{\prime
}}}Q_{k_1k_2}^{k_1^{^{\prime }}k_2^{^{\prime }}}=Q_{k_1k_2}.
\end{equation}
$Q_{k_1k_2}$ is the electric charge operator in fractional-charge
scheme. Therefore, the matrix elements of electric currents should be the same for both schemes of interger charge and fractional charge of quarks.

By using Eqs.(2,3,5,8) and the method of Ref. [1], the
matrix elements of electric currents of $\frac 12^{+}$ baryon are obtained
\begin{eqnarray}
&<&B_{\lambda ^{^{\prime }}}^{\frac 12}(p_f)_{l_1}^{l_1^{^{\prime }}}\mid
J_\mu (0)\mid B_\lambda ^{\frac 12}(p_i)_{l_2}^{l_2^{^{\prime }}}>=-\frac{ie}%
2M^2Q_{k_1k_1^{^{\prime }}}(\gamma _\mu +\frac \kappa {2m_p}q_\nu \sigma
_{\mu \nu })_{\gamma \gamma ^{^{\prime }}}  \nonumber \\
&&\times \smallint d^4x_1d^4x_2\overline{B}_{\alpha \beta \gamma
,ijk_1}^{\frac 12\lambda ^{^{\prime }}}(x_1,x_2,0)_{l_1}^{l_1^{^{\prime
}}}B_{\gamma ^{^{\prime }}\beta \alpha ,k_1^{^{\prime }}ji}^{\frac 12\lambda
}(0,x_2,x_1)_{l_2}^{l_2^{^{\prime }}}  \nonumber \\
&=&-\frac{ie}{24}(\frac{mm^{^{\prime }}}{EE^{^{\prime }}})^{\frac
12}\{A_1I_1-A_2I_2\},
\end{eqnarray}
where M is the rest mass of the quark, m, m' and E, E' are the
initial and final mass and energy of the baryon respectively,
\begin{equation}
A_1=S_p\overline{B}QB,A_2=S_p\overline{B}BQ.
\end{equation}
B, $\overline{B}$ are the $SU_3$ matrices of the initial and
final baryon,
\begin{eqnarray}
I_1 &=&-20\{D_2(q^2)(1-\frac{m_{+}}{5m})+D_2^{^{\prime }}(q^2)(1-\frac{m_{+}%
}{5m^{^{\prime }}})  \nonumber \\
&&+\frac 1{2mm^{^{\prime }}}(m_{-}^2+q^2+\frac{\kappa m_{+}}{5m_p}%
q^2)D_3(q^2)\}\overline{u}_{\lambda ^{^{\prime }}}(p_f)\gamma _\mu u_\lambda
(p_i)  \nonumber \\
&&-20\{2D_1(q^2)-(1-\frac{2m_p}{5\kappa m})D_2(q^2)-(1-\frac{2m_p}{5\kappa
m^{^{\prime }}})D_2^{^{\prime }}(q^2)  \nonumber \\
&&+\frac 1{2mm^{^{\prime }}}(m_{+}^2+\frac 35q^2)D_3(q^2)\}\frac \kappa
{2m_p}\overline{u}_{\lambda ^{^{\prime }}}(p_f)q_\nu \sigma _{\mu \nu
}u_\lambda (p_i)  \nonumber \\
&&-4i\{\frac 1mD_2(q^2)-\frac 1{m^{^{\prime }}}D_2^{^{\prime }}(q^2)+\frac
\kappa {2mm^{^{\prime }}m_p}(m^{^{\prime }2}-m^2)D_3(q^2)\}  \nonumber \\
&&\times q_\mu \overline{u}_{\lambda ^{^{\prime }}}(p_f)u_\lambda (p_i),
\nonumber \\
I_2 &=&4\{(1-2\frac{m_{+}}m)D_2(q^2)+(1-2\frac{m_{+}}{m^{^{\prime }}}%
)D_2^{^{\prime }}(q^2)  \nonumber \\
&&+\frac 1{2mm^{^{\prime }}}(m_{-}^2+q^2+2\frac{\kappa m_{+}}{m_p}%
q^2)D_3(q^2)\}\overline{u}_{\lambda ^{^{\prime }}}(p_f)\gamma _\mu u_\lambda
(p_i)  \nonumber \\
&&-4\{2D_1(q^2)-(1-\frac{4m_p}{\kappa m})D_2(q^2)-(1-\frac{4m_p}{\kappa
m^{^{\prime }}})D_2^{^{\prime }}(q^2)  \nonumber \\
&&+\frac 1{2mm^{^{\prime }}}(m_{+}^2-3q^2)D_3(q^2)\}\frac \kappa {2m_p}%
\overline{u}_{\lambda ^{^{\prime }}}(p_f)q_\nu \sigma _{\nu \mu }u_\lambda
(p_i)  \nonumber \\
&&+8i\{\frac 1mD_2(q^2)-\frac 1{m^{^{\prime }}}D_2^{^{\prime }}(q^2)+\frac{%
\kappa (m^{^{\prime }2}-m^2)}{2mm^{^{\prime }}m_p}D_3(q^2)\}  \nonumber \\
&&\times q_\mu \overline{u}_{\lambda ^{^{\prime }}}(p_f)u_\lambda (p_i),
\end{eqnarray}
where
\begin{equation}
q_\mu =p_{i\mu }-p_{f\mu },m_{+}=m+m^{^{\prime }},m_{-}=m^{^{\prime }}-m,
\end{equation}
\begin{eqnarray}
D_1(q^2) &=&-M^2\smallint f_1^{^{\prime
}}(-x_1,-x_2,0)f_1(0,x_2,x_1)d^4x_1d^4x_2,  \nonumber \\
D_2(q^2) &=&-M^2\smallint f_1^{^{\prime
}}(-x_1,-x_2,0)f_2(0,x_2,x_1)d^4x_1d^4x_2,  \nonumber \\
D_2^{^{\prime }}(q^2) &=&-M^2\smallint f_2^{^{\prime
}}(-x_1,-x_2,0)f_1(0,x_2,x_1)d^4x_1d^4x_2,  \nonumber \\
D_3(q^2) &=&-M^2\smallint f_2^{^{\prime
}}(-x_1,-x_2,0)f_2(0,x_2,x_1)d^4x_1d^4x_2.
\end{eqnarray}
$m, m^{^{\prime }}$ are the rest mass of the initial and final
baryon. $f_j^{^{\prime }}(-x_1,-x_2,0)$, $f_j(0,x_2,x_1)$ are the
spacial  part of the initial and final wave function respectively.
 Eq. (13) shows that when $p_{f\mu
}\longleftrightarrow p_{i\mu }$ is taken, we have
\begin{equation}
D_2(q^2)\longleftrightarrow D_2^{^{\prime }}(q^2),
\end{equation}
therefore when m = m'
\begin{equation}
D_2(q^2)=D_2^{^{\prime }}(q^2).
\end{equation}
Similarly, the current matrix elements of $%
\frac 12^{+}$ baryon$ ->\frac 23^{+}$ baryon are obtained
\begin{eqnarray}
<B_{\lambda ^{^{\prime }}}^{\frac 32}(p_f)^{lmn}\mid J_\mu (0)\mid
B_\lambda ^{\frac 12}(p_i)_{l_1}^{l_1^{^{\prime }}}>
=-\frac{ie}2M^2Q_{kk^{^{\prime }}}(\gamma _\mu +\frac \kappa
{2m_p}q_\nu \sigma _{\mu \nu })_{\gamma \gamma ^{^{\prime }}}
\nonumber \\ \times \smallint d^4x_1d^4x_2\overline{B}_{\alpha
\beta \gamma ,ijk}^{\frac 32\lambda ^{^{\prime
}},lmn}(x_1,x_2,0)B_{\gamma ^{^{\prime }}\beta \alpha ,k^{^{\prime
}}ji}^{\frac 12\lambda }(0,x_2,x_1)_{l_1}^{l_1^{^{\prime }}}
\nonumber
\end{eqnarray}
\begin{eqnarray}
&=&\frac{ie}4(\frac{mm^{^{\prime }}}{EE^{^{\prime }}})d_{l_1jk}^{lmn}%
\varepsilon _{jk^{^{\prime }}l_1^{^{\prime }}}Q_{kk^{^{\prime
}}}\{2D_2(q^2)+\kappa [\frac{m_{+}}{m_p}D_3(q^2)+2\frac m{m_p}D_1(q^2)
\nonumber \\
&&-\frac m{m_p}D_2(q^2)-\frac m{m_p}D_2^{^{\prime }}(q^2)]\}\frac
1{mm^{^{\prime }}}p_\rho q_\sigma \varepsilon _{\rho \sigma \nu \mu }%
\overline{\psi }_\nu ^{\lambda ^{^{\prime }}}(p_f)u_\lambda (p_i)  \nonumber
\\
&&+ie(\frac{mm^{^{\prime }}}{EE^{^{\prime }}})^{\frac
12}d_{l_1jk}^{lmn}\varepsilon _{jk^{^{\prime }}l_1^{^{\prime
}}}Q_{kk^{^{\prime }}}\{D_3(q^2)-D_2(q^2)+\frac{\kappa m}{2m_p}[D_2(q^2)
\nonumber \\
&&+D_2^{^{\prime }}(q^2)-2D_1(q^2)]\}\frac 1{mm^{^{\prime }}}(p_{f\mu }q_\nu
-p_f\cdot q\delta _{\mu \nu })\overline{\psi }_\nu ^{\lambda ^{^{\prime
}}}(p_f)\gamma _5u_\lambda (p_i)  \nonumber \\
&&+ie(\frac{mm^{^{\prime }}}{EE^{^{\prime }}})^{\frac
12}d_{l_1jk}^{lmn}\varepsilon _{jk^{^{\prime }}l_1^{^{\prime
}}}Q_{kk^{^{\prime }}}\{D_2^{^{\prime }}(q^2)-\frac{m^{^{\prime }}}mD_2(q^2)+%
\frac{m_{-}}mD_3(q^2)\}  \nonumber \\
&&\times \overline{\psi }_\mu ^{\lambda ^{^{\prime }}}(p_f)\gamma
_5u_\lambda (p_i).
\end{eqnarray}
m, m' are the rest mass of $\frac 12^{+}$ baryon and $\frac
23^{+}$ baryon respectively,
\begin{equation}
p_\mu =p_{i\mu }+p_{f\mu }.
\end{equation}

In Eq.(11), when m = m' is taken, the terms in $I_1$ and $I_2$, which are
proportional to $q_\mu $ vanish. Thus, when m = m', the current
matrix element of $\frac 12^{+}$ baryon automatically satisfies
current conservation. In general, in order to satisfy
current conservation, the following
relationship must be satisfied
\begin{equation}
D_2^{^{\prime }}(q^2)-\frac{m^{^{\prime }}}mD_2(q^2)+\frac{m_{-}}mD_3(q^2)=0.
\end{equation}
For ${1\over2}^+$ baryons the only matrix element with $m'\neq m$ is $\Sigma
^0->\Lambda $. For this process, we have
\begin{equation}
A_1=A_2=\frac 1{2\sqrt{3}}.
\end{equation}
The condition(18) guarantees current
conservation.

\section{Relationship Between $f_1(x_1,x_2,x_3)$ and $f_2(x_1,x_2,x_3)$}
In this section,
we study the behavior of two invariant
spacial functions $f_1(x_1,x_2,x_3)$ and $f_2(x_1,x_2,x_3)$
in the frame of center-of-mass.
$\Gamma _{\alpha \beta ,\gamma }(p)_\lambda $(4)
can be written as
\begin{eqnarray}
\Gamma _{\alpha \beta ,\gamma }(x_1,x_2,x_3)_\lambda
&=&g_1(x_1,x_2,x_3)\{(1+\gamma _4)\gamma _5C\}_{\alpha \beta
}u_{\lambda ,\gamma }  \nonumber \\
&&+g_2(x_1,x_2,x_3)\{[(1-\gamma _4)\gamma _5C]_{\alpha \beta
}u_{\lambda ,\gamma }+2C_{\alpha \beta }(\gamma _5u_\lambda
)_\gamma \},  \nonumber \\ g_1(x_1,x_2,x_3) &=&\frac
12\{f_1(x_1,x_2,x_3)+f_2(x_1,x_2,x_3)\},  \nonumber \\
g_2(x_1,x_2,x_3) &=&\frac 12\{f_1(x_1,x_2,x_3)-f_2(x_1,x_2,x_3)\},
\end{eqnarray}
$x_1, x_2, x_3$ are the time-space coordinates of three quarks. In
Ref.[1], in order to guarantee $SU_6$ symmetry, it is assumed
that strong interaction satisfies $SU_6$
symmetry when the speed of the quark is much less than the speed
of light. One possibility is that strong
interaction takes scalar form.The B-S equation of
the baryon is written as
\begin{eqnarray}
&&(i\widehat{p}_1+M)_{\alpha \alpha ^{^{\prime }}}(i\widehat{p}_2+M)_{\beta
\beta ^{^{\prime }}}(i\widehat{p}_3+M)_{\gamma \gamma ^{^{\prime
}}}B_{\alpha ^{^{\prime }}\beta ^{^{\prime }}\gamma ^{^{\prime
}},ijk}^{\frac 12\lambda }(p_1,p_2,p_3)_l^m  \nonumber \\
&=&-i(i\widehat{p}_3+M)_{\gamma \gamma ^{^{\prime }}}\smallint U(q)B_{\alpha
\beta \gamma ^{^{\prime }},ijk}^{\frac 12\lambda }(p_1-q,p_2+q,p_3)_l^md^4q
\nonumber \\
&&-i(i\widehat{p}_1+M)_{\alpha \alpha ^{^{\prime }}}\smallint U(q)B_{\alpha
^{^{\prime }}\beta \gamma ,ijk}^{\frac 12\lambda }(p_1,p_2-q,p_3+q)_l^md^4q
\nonumber \\
&&-i(i\widehat{p}_2+M)_{\beta \beta ^{^{\prime }}}\smallint U(q)B_{\alpha
\beta ^{^{\prime }}\gamma ,ijk}^{\frac 12\lambda }(p_1+q,p_2,p_3-q)_l^md^4q
\nonumber \\
&&-\smallint V(q_1,q_2,q_3)\delta ^4(q_1+q_2+q_3)B_{\alpha \beta \gamma
,ijk}^{\frac 12\lambda }(p_1+q_1,p_2+q_2,p_3+q_3)  \nonumber \\
&&\times d^4q_1d^4q_2d^4q_3.
\end{eqnarray}
$B_{\alpha \beta \gamma ,ijk}^{\frac 12\lambda }(p_1,p_2,p_3)_l^m$
is the wave function for $\frac 12^{+}$ baryon in the frame of
center-of-mass. We assume $U(q)$
and $V(q_1,q_2,q_3)$ are independent of the momentum of
the baryon.
\begin{equation}
p_1+p_2+p_3=p,
\end{equation}
where p is the momentum of $\frac 12^{+}$ baryon.

According to Ref.[2], in
order to satisfy
$SU_6$ symmetry
the terms at $O(\frac{|{\bf
p}_j|}M(j=1,2,3))$ in the wave function are ignored. The same
treatment is used in Eq.(21). Substituting the wave function
of $\frac 12^{+}$ into
Eq.(21), we obtain
\begin{eqnarray}
&&(M-\gamma _4p_{10})_{\alpha \alpha ^{^{\prime }}}(M-\gamma
_4p_{20})_{\beta \beta ^{^{\prime }}}(M-\gamma _4p_{30})_{\gamma \gamma
^{^{\prime }}}\Gamma _{\alpha ^{^{\prime }}\beta ^{^{\prime }}\gamma
^{^{\prime }}}(p_1,p_2,p_3)_\lambda   \nonumber \\
&=&-i(M-\gamma _4p_{30})_{\gamma \gamma ^{^{\prime }}}\smallint U(q)\Gamma
_{\alpha \beta ,\gamma ^{^{\prime }}}(p_1-q,p_2+q,p_3)_\lambda d^4q
\nonumber \\
&&-i(M-\gamma _4p_{10})_{\alpha \alpha ^{^{\prime }}}\smallint U(q)\Gamma
_{\alpha ^{^{\prime }}\beta ,\gamma }(p_1,p_2-q,p_3+q)_\lambda d^4q
\nonumber \\
&&-i(M-\gamma _4p_{20})_{\beta \beta ^{^{\prime }}}\smallint U(q)\Gamma
_{\alpha \beta ^{^{\prime }},\gamma }(p_1+q,p_2,p_3-q)_\lambda d^4q
\nonumber \\
&&-\smallint V(q_1,q_2,q_3)\delta ^4(q_1+q_2+q_3)\Gamma _{\alpha \beta
,\gamma }(p_1+q_1,p_2+q_2,p_3+q_3)  \nonumber \\
&&\times d^4q_1d^4q_2d^4q_3.
\end{eqnarray}
where $\Gamma _{\alpha \beta, \gamma }(p_1,p_2,p_3)_\lambda $ is
the expression of $\Gamma _{\alpha \beta \gamma
}(x_1,x_2,x_3)_\lambda $(20) in the momentum
representation. Calculations lead to
\begin{eqnarray}
&&(M-p_{10})(M-p_{20})(M-p_{30})g_1(p_1,p_2,p_3)  \nonumber \\
&=&-i\smallint U(q)\{(M-p_{30})g_1(p_1-q,p_2+q,p_3)  \nonumber \\
&&+(M-p_{10})g_1(p_1,p_2-q,p_3+q)  \nonumber \\
&&+(M-p_{20})g_1(p_1+q,p_2,p_3-q)\}d^4q  \nonumber \\
&&-\smallint V(q_1,q_2,q_3)\delta ^4(q_1+q_2+q_3)  \nonumber \\
&&\times g_1(p_1+q_1,p_2+q_2,p_3+q_3)d^4q_1d^4q_2d^4q_3,
\end{eqnarray}
\begin{eqnarray}
&&(M+p_{10})(M+p_{20})(M-p_{30})g_2(p_1,p_2,p_3)  \nonumber \\
&=&-i\smallint U(q)\{(M-p_{30})g_2(p_1-q,p_2+q,p_3)  \nonumber \\
&&+(M+p_{10})g_2(p_1,p_2-q,p_3+q)  \nonumber \\
&&+(M+p_{20})g_2(p_1+q,p_2,p_3-q)\}d^4q  \nonumber \\
&&-\smallint V(q_1,q_2,q_3)\delta ^4(q_1+q_2+q_3)  \nonumber \\
&&\times g_2(p_1+q_1,p_2+q_2,p_3+q_3)d^4q_1d^4q_2d^4q_3,
\end{eqnarray}
\begin{eqnarray}
&&(M+p_{10})(M-p_{20})(M+p_{30})g_2(p_1,p_2,p_3)  \nonumber \\
&=&-i\smallint U(q)\{(M+p_{30})g_2(p_1-q,p_2+q,p_3)  \nonumber \\
&&+(M+p_{10})g_2(p_1,p_2-q,p_3+q)  \nonumber \\
&&+(M-p_{20})g_2(p_1+q,p_2,p_3-q)\}d^4q  \nonumber \\
&&-\smallint V(q_1,q_2,q_3)\delta ^4(q_1+q_2+q_3)  \nonumber \\
&&\times g_2(p_1+q_1,p_2+q_2,p_3+q_3)d^4q_1d^4q_2d^4q_3,
\end{eqnarray}
\begin{eqnarray}
&&(M-p_{10})(M+p_{20})(M+p_{30})g_2(p_1,p_2,p_3)  \nonumber \\
&=&-i\smallint U(q)\{(M+p_{30})g_2(p_1-q,p_2+q,p_3)  \nonumber \\
&&+(M-p_{10})g_2(p_1,p_2-q,p_3+q)  \nonumber \\
&&+(M+p_{20})g_2(p_1+q,p_2,p_3-q)\}d^4q  \nonumber \\
&&-\smallint V(q_1,q_2,q_3)\delta ^4(q_1+q_2+q_3)  \nonumber \\
&&\times g_2(p_1+q_1,p_2+q_2,p_3+q_3)d^4q_1d^4q_2d^4q_3.
\end{eqnarray}
Since $V(q_1,q_2,q_3)$ are totally symmetric functions of $q_1,
q_2, q_3$. $g_1(p_1,p_2,p_3)$ are
totally symmetric functions of $p_1, p_2, p_3$, which is
consistent with Ref.[2]. From Eqs.(25-27), we see that
$g_2(p_1,p_2,p_3)$ have following symmetries: (1) totally
symmetric in $p_1, p_2, p_3$. (2) since $U(q)$ and
$V(q_1,q_2,q_3)$ are independent of the momentum p
, the equation is invariant under the transformations
$p_{20}\rightarrow -p_{20}$, $p_{30}\rightarrow -p_{30}$;
$p_{10}\rightarrow -p_{10}$, $p_{30}\rightarrow -p_{30}$;
$p_{10}\rightarrow -p_{10}$, $p_{20}\rightarrow -p_{20}$. By
using the second symmetry of $g_2(p_1,p_2,p_3)$, Eq.(3.6)
becomes Eq.(3.5) under the transformation $p_{10}\rightarrow
-p_{10}$, $p_{20}\rightarrow -p_{20}$, thus $g_1(p_1,p_2,p_3)$ and
$g_2(p_1,p_2,p_3)$ satisfy the same equation. $g_1(p_1,p_2,p_3)$
is related to $g_2(p_1,p_2,p_3)$ by
\begin{equation}
g_1(p_1,p_2,p_3)=bg_2(p_1,p_2,p_3),
\end{equation}
where b is a constant. Eq.(28) leads to
\begin{equation}
f_2(x_1,x_2,x_3)=af_1(x_1,x_2,x_3).
\end{equation}
Thus, in the wave functions (3,5), there is only
one independent spacial function.

Substituting Eq.(29) into
Eq.(18), we obtain
\begin{equation}
a=\frac 1{1-\frac{m_0}m} \hspace*{.1in} or \hspace*{.1in} a=1.
\end{equation}
$m_0$ is a parameter, m is the physical mass of the baryon.
Generally $a\neq 1$, a takes the first expression of
Eq.(30).

\section{Electromagetice Properties of $\frac 12^{+}$ Baryons}
The electromagnetic form factors of $\frac 12^{+}$ baryon are
obtained from the current matrix elements Eq.(9,11)
\begin{eqnarray}
G_E(q^2) &=&-\frac 23(A_1+2A_2)(1+\frac{q^2}{4m^2})\{D_2(q^2)-\frac{\kappa
q^2}{4mm_p}D_3(q^2)\}  \nonumber \\
&&+\frac 13(A_2+5A_1)\{D_2(q^2)+\frac{q^2}{4m^2}[D_3(q^2)+\frac{\kappa m}{m_p%
}D_2(q^2)  \nonumber \\
&&-\frac{\kappa m}{m_p}D_1(q^2)-\kappa \frac m{m_p}(1+\frac{q^2}{4m^2}%
)D_3(q^2)]\}. \\
G_M(q^2) &=&\frac 13(A_2+5A_1)\{D_2(q^2)+\frac{q^2}{4m^2}D_3(q^2)+\kappa
\frac m{m_p}[D_1(q^2)  \nonumber \\
&&-D_2(q^2)+(1+\frac{q^2}{4m^2})D_3(q^2)]\},
\end{eqnarray}
where m is the mass of the baryon. From Eq.(31) we obtain
\begin{equation}
D_2(0)=1.
\end{equation}
The expression of the magnetic moment of $\frac 12^{+}$ baryon is
obtained from Eq.(32)
\begin{eqnarray}
\mu  &=&\frac 13(A_2+5A_1)\{\frac{m_p}m+\kappa
[D_1(0)+D_3(0)-1]\}.
\end{eqnarray}
From Eqs.(40,41,33)
\begin{eqnarray}
\mu  &=&\frac 13(A_2+5A_1)\{\frac{m_p}m+\kappa (\frac 1{1-\frac{m_0}m}-\frac{%
m_0}m)\}
\end{eqnarray}
is obtained.
The two parameters $\kappa, m_0$ in Eq.(35) are
determined to be
\begin{equation}
\kappa =0.481,m_0=0.778m_p
\end{equation}
by input the magnetic moments of proton and $\Sigma $
[8]. The magnetic moments of other
six baryons are determined to be
\begin{center}
\begin{tabular}{c|c|c|c|c|c|c|c|c} \hline
& $\mu _p$ & $\mu _n$ & $\mu _\Lambda $ & $\mu _{\Sigma ^{+}}$ &
$\mu _{\Sigma ^{0}}$ & $\mu _{\Sigma ^{-}}$ & $\mu _{\Xi ^0}$ &
$\mu _{\Xi ^-}$ \\ \hline theory & 2.79 &-.1.86 & -0.64 & 1.74 &
0.58 & -0.57 & -0.97 & -0.51 \\& (input) & & (input) & & &&  & \\
\hline exp & 2.79 & -1.91 & -0.67 &2.59& & & & -1.93 \\ & && $\pm
0.46$ & $\pm 0.46$ & & & & $\pm 0.75$ \\ \hline
\end{tabular}
\end{center}
The electromagnetic form factors of proton and neutron are found
from Eq.(31,32)
\begin{eqnarray}
G_E^p(q^2) &=&D_2(q^2)+\frac{q^2}{4m_N^2}\{D_3(q^2)+\kappa
[D_2(q^2)-D_1(q^2)-(1+\frac{q^2}{4m_N^2})D_3(q^2)]\},  \nonumber \\
G_M^p(q^2) &=&D_2(q^2)+\kappa [D_1(q^2)+D_3(q^2)-D_2(q^2)]+(1+\kappa )\frac{%
q^2}{4m_N^2}D_3(q^2),  \\ G_E^n(q^2) &=&-\frac
23\frac{q^2}{4m_N^2}\{D_3(q^2)-D_2(q^2)+\kappa
[D_2(q^2)-D_1(q^2)]\},  \nonumber \\ G_M^n(q^2) &=&-\frac
23G_M^p(q^2).
\end{eqnarray}
By using Eqs.(29,30,36)
\begin{eqnarray}
G_E^p(q^2) &=&D_2(q^2)\{1+\tau (2.71-2.17\tau )\},  \nonumber \\
G_M^p(q^2) &=&\mu _pD_2(q^2)\{1+2.39\tau \},  \nonumber \\
G_E^n(q^2) &=&1.39\mu _n\tau D_2(q^2)
\end{eqnarray}
are obtained, where \(\tau =\frac{q^2}{4m_N^2}\).
It is seen from Eq.(39) that there is an invariant function
D$_2(q^2)$ in the three form factors, which can be determined
from the experimental data of the magnetic form factor of
proton [9]
\begin{equation}
D_2(q^2)=\frac 1{(1+\frac{q^2}{0.71})^2(1+2.39\tau )}.
\end{equation}
The ratio of the electric and
magnetic form factor of proton is obtained
\begin{equation}
\frac{\mu _pG_E^p(q^2)}{G_M^p(q^2)}=\frac{1+\tau (2.71-2.17\tau )}{%
1+2.39\tau }.
\end{equation}
Comparisons with data are shown in Fig.1 and 2.
The experimental data of Fig.1 is from Ref.[10], and that for
Fig.2 is from Ref.[11].

The expression of the electric form factor of
neutron is obtained
\begin{equation}
G_E^n(q^2)=1.39\tau G_M^n(q^2) (1+2.39\tau )^{-1}.
\end{equation}
The slope of $G_E^n(q^2)$ at $q^2=0$ is
\begin{equation}
\frac{dG_E^n(q^2)}{dq^2}\mid _{q^2=0}=1.39\frac{\mu _n}{4m_N^2}%
=-0.73[GeV]^{-2}.
\end{equation}
The experimental data are
\begin{equation}
-0.579\pm 0.018^{[12]},-0.512\pm 0.049^{[13]},0.495\pm 0.010^{[14]}.
\end{equation}
Comparisons of Eq.(42) with the experimental data are shown in
Fig.3 and Fig.4.
The experimental data of Fig.3 comes from Ref.[4] and that for
Fig.4 comes from Ref.[11].

At $q^2=-4m_N^2$, there are
\begin{eqnarray}
G_E^p(-4m^2) &=&G_M^p(-4m^2)=0.18,  \nonumber \\
G_E^n(-4m^2) &=&G_M^p(-4m^2)=-\frac 23G_E^p(-4m^2).
\end{eqnarray}

The S-matrix element of $\Sigma ^0\rightarrow \Lambda
+\gamma $ is studied
\begin{eqnarray}
<\gamma \Lambda \mid S\mid \Sigma ^0>&=&-ie(2\pi )^4\delta
(p_i-p_f-q)\frac{e_\mu ^\lambda }{\sqrt{2\omega }}(\frac{m_\Lambda
}{E_\Lambda })^{\frac 12}\mu _{\Sigma _0\Lambda }  \nonumber \\
&&\times \frac \kappa {2m_p}\overline{u}_{\lambda ^{^{\prime
}}}(p_f)q_\nu \sigma _{\nu \mu }u_\lambda (p_i),
\end{eqnarray}
\begin{eqnarray}
 \mu _{\Sigma _0\Lambda } &=&\frac
1{2\sqrt{3}}D_3(0)\{\frac 2{a_\Lambda a_{\Sigma ^0}}-\frac
1{a_\Lambda }(1-\frac{m_p}{\kappa m_\Sigma })-\frac 1{a_{\Sigma
^0}}(1-\frac{m_p}{\kappa m_\lambda })+\frac{m_{+}^2}{2m_\Lambda
m_\Sigma }\}.
\end{eqnarray}
The dependences of $D_1(0)$, $D_2(0)$,
$D_1^{^{\prime }}(0)$ and $D_1^{^{\prime }}(0)$ on the mass of
initial and final baryon need to be found. From Eq.(28) we have
\begin{equation}
\frac{D_2(0)}{D_2^{^{\prime }}(0)}=\frac a{a^{^{\prime }}}.
\end{equation}
On the other hand,
Eq.(14) shows when $m\longleftrightarrow m^{^{\prime
}}$ is taken, we have
\begin{eqnarray}
D_2(0)& \longleftrightarrow &D_2^{^{\prime }}(0).
\end{eqnarray}
When m = m', Eqs.(15,33) lead to
\begin{eqnarray}
 D_2(0) =D_2^{^{\prime }}(0)=1.
\end{eqnarray}
The general expressions of $D_2(0)$, $D_2^{^{\prime }}(0)$ which satisfy
Eqs.(48-50) are found
\begin{eqnarray}
D_2(0) &=&(\frac a{a^{^{\prime }}})^{\frac 12}f(m,m^{^{\prime }}),  \nonumber
\\
D_2^{^{\prime }}(0) &=&(\frac{a^{^{\prime }}}a)^{\frac 12}f(m,m^{^{\prime
}}).
\end{eqnarray}
$f(m,m^{^{\prime }})$ is a symmetric function of m, m' and
\begin{equation}
f(m,m)=1.
\end{equation}
When $m\neq m^{^{\prime }}$, the deviation of $f(m,m^{^{\prime
}})$ from 1 is proportional to $(m-m^{^{\prime }})^2$. According
to Ref.[1], $f(m,m^{^{\prime }})$ is the effect of  Lorentz contraction. We
obtain
\begin{equation}
f(m,m^{^{\prime }})=\frac{4mm^{^{\prime }}}{(m+m^{^{\prime }})^2}.
\end{equation}
For $\Sigma ^0\rightarrow \Lambda +\gamma $, the deviation of
$f(m,m^{^{\prime }})$  from 1 is only $0.1\%$. From Eq.(51), the
expressions of $D_1(0)$ and $D_3(0)$ are found
\begin{eqnarray}
D_3(0) &=&\sqrt{aa^{^{\prime }}}f(m,m^{^{\prime }}),  \nonumber \\
D_1(0) &=&\frac 1{\sqrt{aa^{^{\prime }}}}f(m,m^{^{\prime }}).
\end{eqnarray}
The magnetic moment of $\Sigma\rightarrow\Lambda$
and the decay rate are computed to be
\begin{equation}
\mu _{\Sigma _0\Lambda }=1.053
\end{equation}
\begin{eqnarray}
\Gamma  &=&\frac \alpha 8\mu _{\Sigma ^0\Lambda }^2\frac{m_\Sigma ^3}{m_p^2}%
(1-\frac{m_\Lambda ^2}{m_\Sigma ^2})^3=3.79\times 10^{-3}MeV,  \nonumber \\
\tau  &=&\frac 1\Gamma =1.74\times 10^{-19}sec.
\end{eqnarray}
The experimental upper limit is
\begin{equation}
\tau <1.0\times 10^{-14}sec. \nonumber
\end{equation}

\section{Electromagnetic transition of
$p\rightarrow \Delta (1236)$}

The matrix elements of currents are obtained from Eqs.(18,28,16).
Substituting Eqs.(18),(28) into Eq.(16), we derive
\begin{eqnarray}
&<&\Delta _{\lambda ^{^{\prime }}}^{+}(p_f)\mid J_\mu (0)\mid p_\lambda
(p_i)>=-\frac{ie}{4\sqrt{3}}(\frac{mm^{^{\prime }}}{EE^{^{\prime }}})^{\frac
12}A\frac 1{mm^{^{\prime }}}D_3(q^2)p_\rho q_\sigma   \nonumber \\
&&\times \varepsilon _{\rho \sigma \nu \mu }\overline{\psi }_\nu ^{\lambda
^{^{\prime }}}(p_f)u_\lambda (p_i)-\frac{ie}{\sqrt{3}}(\frac{mm^{^{\prime }}%
}{EE^{^{\prime }}})^{\frac 12}\frac B{mm^{^{\prime }}}D_3(q^2)  \nonumber \\
&&\times (p_{f\mu }q_\nu -p_f\cdot q\delta _{\mu \nu })\overline{\psi }_\nu
^{\lambda ^{^{\prime }}}(p_f)\gamma _5u_\lambda (p_i),
\end{eqnarray}
where
\begin{eqnarray}
A &=&\frac 2{a^{^{\prime }}}+\kappa \{1+\frac{m^{^{\prime }}}{m_p}+\frac
2{aa^{^{\prime }}}-\frac 1a-\frac 1{a^{^{\prime }}}\},  \nonumber \\
B &=&1-\frac 1{a^{^{\prime }}}+\frac \kappa 2\{\frac 1a+\frac 1{a^{^{\prime
}}}-\frac 2{aa^{^{\prime }}}\},
\end{eqnarray}
and
\begin{equation}
A=1.717,B=0.699.
\end{equation}

The S matrix element of $\gamma p\rightarrow\pi N$ is written as
\begin{eqnarray}
<\pi N\mid S\mid \gamma p>&=&-i(2\pi )^4\delta (p_\gamma
+p_i-p_\pi -p_N)\sum_{\lambda ^{^{\prime }}}<\pi N\mid U\mid
\Delta _{\lambda ^{^{\prime }}}^{+}(p_f)>  \nonumber \\ &&\times
<\Delta _{\lambda ^{^{\prime }}}^{+}(p_f)\mid U\mid \gamma
p>\frac{E_\Delta }{m_\Delta }\frac 1{W-m_\Delta +\frac i2\Gamma
(W)},
\end{eqnarray}
where W is the mass of the final state, $\Gamma (W)$ is the total
width of the strong decay of $\Delta (1236)$. The calculation is
done in the rest frame of $\Delta (1236)$. $<\pi N\mid U\mid
\Delta _{\lambda ^{^{\prime }}}^{+}(p_f)>$ is the amplitude of the
strong decay of $\Delta (1236)$
\begin{eqnarray}
<\pi N\mid U\mid \Delta _{\lambda ^{^{\prime }}}^{+}(p_f)>=(\frac{m_N}{%
2E_\pi E_N})^{\frac 12}g(W)\frac{p_{\pi \mu }}{m_N}\overline{u}(p_N)\psi
_\mu ^{\lambda ^{^{\prime }}}.
\end{eqnarray}
The electric transition amplitude in Eq.(60) is expressed as
\begin{eqnarray}
<\Delta _{\lambda ^{^{\prime }}}\mid U\mid \gamma p>=-\frac 1{\sqrt{%
2E_\gamma }}e_\mu <\Delta _{\lambda ^{^{\prime }}}\mid J_\mu (0)\mid p>.
\end{eqnarray}
By using following equation
\begin{eqnarray}
\sum_{\lambda ^{^{\prime }}}\psi _\mu ^{\lambda ^{^{\prime }}}\overline{\psi
}_{\mu ^{^{\prime }}}^{\lambda ^{^{\prime }}} &=&\frac 13(1+\gamma
_4)\{\delta _{\mu \mu ^{^{\prime }}}+\frac 12\gamma _5\gamma _j\varepsilon
_{j\mu \mu ^{^{\prime }}}\}  \nonumber \\
(j,\mu ,\mu ^{^{\prime }} &=&1,2,3)
\end{eqnarray}
and Eq.(58), we obtain
\begin{eqnarray}
\sum_{\lambda ^{^{\prime }}}\overline{u}_\gamma (p_N)\psi _\mu ^{\lambda
^{^{\prime }}} &<&\Delta _{\lambda ^{^{\prime }}}\mid J_\nu (0)\mid
p_\lambda >p_{\pi \mu }e_\nu   \nonumber \\
&=&\frac{eD_3(0)}{24\sqrt{3}m_N^2m_\Delta }\{\frac{m_N(m_N+E_N)}{E_i(m_N+E_i)%
}\}^{\frac 12}\overline{u}_\gamma \{[A(m_N+m_\Delta )^2+B(m_\Delta
^2-m_N^2)] \nonumber \\ &&\times [2{\bf k\cdot (e\cdot p_\pi}
)+i{\bf \sigma \cdot ep_\pi \cdot k}-i{\bf \sigma \cdot kp_\pi
\cdot e}] \nonumber \\ &&-3iB(m_\Delta ^2-m_N^2)({\bf \sigma \cdot
ep_\pi \cdot k}+{\bf \sigma \cdot kp_\pi \cdot e})\}u_\lambda ,
\end{eqnarray}
where $E_i$ is the energy of the initial proton, k is the energy
of the photon. The amplitudes of the magnetic and electric transitions
are obtained by comparing with the photo production amplitudes in Ref.[15]
\begin{eqnarray}
M1+ &=&\frac{eD_3(0)}{96\sqrt{3}\pi m_N^2m_\Delta }\{\frac{m_N+E_N}{m_\Delta
E_i(m+E_i)}\}^{\frac 12}\frac{g(W)p_\pi k}{W-m_\Delta +\frac i2\Gamma (W)}
\nonumber \\
&&\times \{A(m_N+m_\Delta )^2+B(m_0^2-m_N^2)\}, \\
E1+ &=&-\frac{eD_3(0)}{96\sqrt{3}\pi m_N^2m_\Delta }\{\frac{m_N+E_N}{%
m_\Delta E_i(m+E_i)}\}^{\frac 12}\frac{g(W)p_\pi k}{W-m_\Delta +\frac
i2\Gamma (W)}  \nonumber \\
&&\times B(m_\Delta ^2-m_N^2),
\end{eqnarray}
\begin{equation}
\frac{E1+}{M1+}=\frac{-B(m_\Delta -m_N)}{A(m_\Delta +m_N)+B(m_\Delta -m_N)}%
=-5.4\%.
\end{equation}
There are several experimental values: -0.045 [16], -0.073 [17],
-0.024 [18].

The partial width of $\Delta
^{+}(1236)\rightarrow p+\gamma $ is derived
\begin{eqnarray}
\Gamma _\gamma  &=&\frac{k^2}{2\pi }\frac{m_N}{m_\Delta }\{\mid
A_{\frac 32}\mid ^2+\mid A_{\frac 12}\mid ^2\},  \\
A_{\frac 32} &=&-\frac{eD_3(0)(m_\Delta +m_N)(m_\Delta ^2-m_N^2)}{8\sqrt{6}%
(m_Nm_\Delta )^{3/2}}\{A+2B\frac{m_\Delta -m_N}{m_\Delta +m_N)}\}  \nonumber
\\
&=&-0.21[GeV]^{-\frac 12},  \nonumber \\
A_{\frac 12} &=&-\frac{eD_3(0)(m_\Delta +m_N)(m_\Delta ^2-m_N^2)}{24\sqrt{2}%
(m_Nm_\Delta )^{3/2}}\{A-2B\frac{m_\Delta -m_N}{m_\Delta +m_N)}\}  \nonumber
\\
&=&-0.10[GeV]^{-\frac 12}.
\end{eqnarray}
The experimental data
are [19]
\begin{equation}
A_{\frac 32}=-0.24[GeV]^{-\frac 12},A_{\frac 12}=-0.14[GeV]^{-\frac 12}.
\end{equation}
The decay width is computed to be
\begin{equation}
 \Gamma _\gamma =0.64MeV,
\end{equation}
The experimental data [21] is 0.65MeV.

\section{Magnetic moment and electromagnetic form factors of
$p\rightarrow \Delta ^{+}(1236)$}

The differential cross section of the
electric production
\begin{center}
$e+p\rightarrow e+\Delta ^{+}(1236)$ \\ \hspace*{0.9in}
$\hookrightarrow N+\pi$
\end{center}
is expressed as
\begin{equation}
\frac 1{\Gamma _t}\frac{d^2\sigma }{d\Omega dE^{^{\prime }}}=\sigma
_T+\varepsilon \sigma _S.
\end{equation}
where E' is the energy of the outgoing electron. Use of
the equation
\begin{eqnarray}
\sum_\lambda \psi _\mu ^\lambda (p)\overline{\psi }_{\mu ^{^{\prime
}}}^\lambda (p) &=&\frac 12(1-\frac i{m^{^{\prime }}}\widehat{p})\{\delta
_{\mu \mu ^{^{\prime }}}+\frac 23\frac{p_\mu p_{\mu ^{^{\prime }}}}{%
m^{^{\prime }2}}-\frac 13\gamma _\mu \gamma _{\mu ^{^{\prime }}}  \nonumber
\\
&&-\frac i{3m^{^{\prime }}}(p_\mu \gamma _{\mu ^{^{\prime }}}-p_{\mu
^{^{\prime }}}\gamma _\mu )\}
\end{eqnarray}
and Eq.(58) leads to
\begin{eqnarray}
\sigma _T &=&\frac{m\alpha q^{*2}}{m^{^{\prime }}(W^2-m^2)}\frac{\Gamma (W)}{%
(W-m^{^{\prime }})^2+\frac 14\Gamma ^2(W)}\frac{D_3^2(q^2)}{18m^2}%
\{A^2(q^2+m_{+}^2)  \nonumber \\
&&+2AB(m^{^{\prime }2}-m^2-q^2)+4B^2(q^2+m_{-}^2)(1-\frac{q^2}{q^{*2}})\}, \\
\sigma _S &=&\frac{m\alpha q^{*2}}{m^{^{\prime }}(W^2-m^2)}\frac{\Gamma (W)}{%
(W-m^{^{\prime }})^2+\frac 14\Gamma ^2(W)}\frac{2D_3^2(q^2)}{9m^2}%
B^2(q^2+m_{+}^2)\frac{q^2}{q^{*2}},
\end{eqnarray}
where
\begin{equation}
 W^2=-(p_i+p_e-p_{e^{^{\prime }}})^2,q^{*2}=q^2+\frac
1{4m^{^{\prime }2}}(m^{^{\prime }2}-m^2-q^2)^2.
\end{equation}
The ratio of $\sigma _S$ and $\sigma _T$ is obtained
\begin{eqnarray}
R &=&\frac{\sigma _S}{\sigma _T}=[4B^2(q^2+m_{-}^2)\frac{q^2}{q^{*2}}%
]/[(Am_{+}+Bm_{-})^2  \nonumber \\
&&+(A-B)^2q^2+3B^2(q^2+m^2)-4B^2(q^2+m_{-}^2)\frac{q^2}{q^{*2}}].
\end{eqnarray}
The behavior of R is released
\begin{eqnarray}
q^2&=&0,\;\;\;R=0;  \nonumber \\
q^2&\rightarrow&\infty,\;\;\;R\sim \frac 1{q^2}\rightarrow 0.
\end{eqnarray}
In the range of $q^2>3[GeV]^2$, $R\sim 0.27$.

According to the definition of mutipoles, the magnetic transition
form factor $G_{M1+}^2(q^2)$, the electric transition form factor
$G_{E1+}^2(q^2)$ and $G_{S1+}^2(q^2)$ are found
\begin{eqnarray}
G_{M1+}^2(q^2) &=&\frac{D_3^2(q^2)}{18m^2}%
\{(Am_{+}+Bm_{-})^2+(A-B)q^2-B^2(q^2+m_{-}^2)\frac{q^2}{q^{*2}}\}, \\
G_{E1+}^2(q^2) &=&\frac{D_3^2(q^2)}{18m^2}B^2(q^2+m_{-}^2)(1-\frac{q^2}{%
q^{*2}}), \\
G_{S1+}^2(q^2) &=&\frac{D_3^2(q^2)}{18m^2}B^2(q^2+m_{-}^2).
\end{eqnarray}
The differential cross
section(73) is expressed as
\begin{eqnarray}
\frac 1{\Gamma _t}\frac{d^2\sigma }{dE^{^{\prime }}d\Omega } &=&\frac{%
m\alpha q^{*2}}{m^{^{\prime }}(W^2-m^2)}\{G_{M1+}^2(q^2)+3G_{E1+}^2(q^2)
\nonumber \\
&&+4\varepsilon G_{S1+}^2(q^2)\frac{q^2}{q^{*2}}\}\frac{\Gamma (W)}{%
(W-m^{^{\prime }})^2+\frac 14\Gamma ^2(W)}.
\end{eqnarray}
From Eq.(79), the magnetic moment of $p\rightarrow
\Delta ^{+}(1236)$ is derived
\begin{equation}
\mu =G_{M1+}(0)=\frac{D_3(0)}{3\sqrt{2}m}(Am_{+}+Bm_{-})^2=1.23\frac{2\sqrt{2%
}}3\mu _p.
\end{equation}
The data are
\begin{equation}
1.22\frac{2\sqrt{2}}3\mu _p^{[18]},\;\;\;1.28\frac{2\sqrt{2}}3\mu
_p^{[21]}.  \nonumber
\end{equation}
Lorentz contraction effect is considered in Eq.(84), for
$p\rightarrow \Delta ^{+}(1236)$
\begin{equation}
f(m,m^{^{\prime }})=0.98.
\end{equation}
If this effect is ignored, the theoretical value of the magnetic
transition moment is $1.26\frac{2\sqrt{2}}3\mu _p$. The electric
multipole moments are obtained from Eq.(81)
\begin{equation}
E1+=-\frac{D_3(0)}{3\sqrt{2}m}Bm_{-}=-0.17,
\end{equation}
\begin{eqnarray}
S1+ &=&E1+,  \nonumber \\
\frac{S1+}\mu  &=&-5.4\%.
\end{eqnarray}
The data [4] is
\begin{equation}
\frac{S1+}{\mu}=(-5\pm3)\%.
\end{equation}
Theoretical results agree with experimental data.

The expression
\begin{equation}
\sigma _T^R=\frac{m\alpha q^{*2}}{m^{^{\prime }}(W^2-m^2)}\frac{\Gamma (W)}{%
(W-m^{^{\prime }})^2+\frac 14\Gamma ^2(W)}G_M^2(q^2)
\end{equation}
has been used to determine $G_M^2(q^2)$. $G_M$ is obtained from Eq.(75) that
\begin{eqnarray}
G_M^2(q^2) &=&G_{M1+}^2(q^2)+3G_{E1+}^2(q^2)  \nonumber \\
&=&\frac{D_3^2(q^2)}{18m^2}%
\{(Am_{+}+Bm_{-})^2+(A-B)^2q^2+B^2(q^2+m_{-}^2)(3-4\frac{q^2}{q^{*2}})\}.
\end{eqnarray}
The mass difference between $\Delta ^{+}(1236)$ and proton is
ignored. The integral $D_3(q^2)$ for $p\rightarrow \Delta
^{+}(1236)$ is expressed as
\begin{equation}
D_3(q^2)=\frac{4mm^{^{\prime }}\sqrt{aa^{^{\prime }}}}{(m+m^{^{\prime }})^2}%
(1+2.39\frac{q^2}{4m^2})^{-1}(1+\frac{q^2}{0.71})^{-1}.
\end{equation}
Substituting Eq.(92) into Eq.(91), the expression of
$G_M^2(q^2)$ is obtained. Comparisons with experimental data
are shown in Fig.5 and 6. The data
for Fig.5 comes from Ref.[4] and that for Fig.6 comes from Ref.[22].
It can be seen from these two figures that as $q^2$ increases, the
theoretical curve drops a little bit faster than the experimental
one. At $q^2=0.8[GeV]^2$, the theoretical value is $10\%$ less
than the experimental value. This difference can be regarded as to
be from the ignorance of the mass difference between proton and
$\Delta ^{+}(1236)$.

Taking
\begin{equation}
W=m^{^{\prime }}=1.236GeV,\;\;\;\Gamma (m^{^{\prime }})=0.12GeV,
\end{equation}
we obtain
\begin{equation}
\sigma _S=48.4q^2(q^2+0.0888)(1+0.679q^2)^{-2}(1+\frac{q^2}{0.71}%
)^{-4}\times 10^{-28}cm^2.
\end{equation}
Comparison with the data [23] is shown in Fig.7.

\section{Discussion}
$SU(6)$ summetric wave functions of ${1\over2}^+$ and ${3\over2}^+$ of s-
wave(in the frame of center of mass) are applied to study the electromagnetic form factors of nucleons and $p\rightarrow\Delta$. A new expression of
$\frac{\mu_p G^p_E(q^2)}{G^p_{M}(q^2)}$ is obtained.
Nonzero electric form factor
$G^n_E(q^2)$ is found. The magnetic form factor of $p\rightarrow\Delta$ decreases faster than $G^p_M$. Nonzero multipole moments $E1+$ and
$S1+$ are obtained. They are small and negative. It is interesting to
point out that nonzero $G^n_E$, $E1+$, and $S1+$ are resulted in the addtional
terms of the wave function, which are constructed by the spinors of antiquarks.
The amplitudes and decay rate of $\Delta\rightarrow p+\gamma$ are computed and theory agrees with data.

The magnetic moments of hyperons are calculated under the assumption that the
anomalous magnetic moment of strange quark(except for the charge factor)
is the same as the one of u and d quarks.
 \pagebreak

\pagebreak
\begin{flushleft}
{\bf Figure Captions}
\end{flushleft}
{\bf FIG. 1.} Ratio of electric and magnetic form factors of proton.
\\ {\bf FIG. 2.}
Ratio of electric and magnetic form factors of proton.
\\{\bf FIG. 3.} Electic form factor of neutron.
\\ {\bf FIG. 4.} Electric form factor of neutron.
\\{\bf FIG. 5.} Magnetic form factor of $p\rightarrow\Delta$.
\\{\bf FIG. 6.} Magnetic form factor of $p\rightarrow\Delta$.
\\{\bf FIG. 7.} Cross Section of virtual scalar photon.

\begin{figure}
\begin{center}
\includegraphics[width=7in, height=7in]{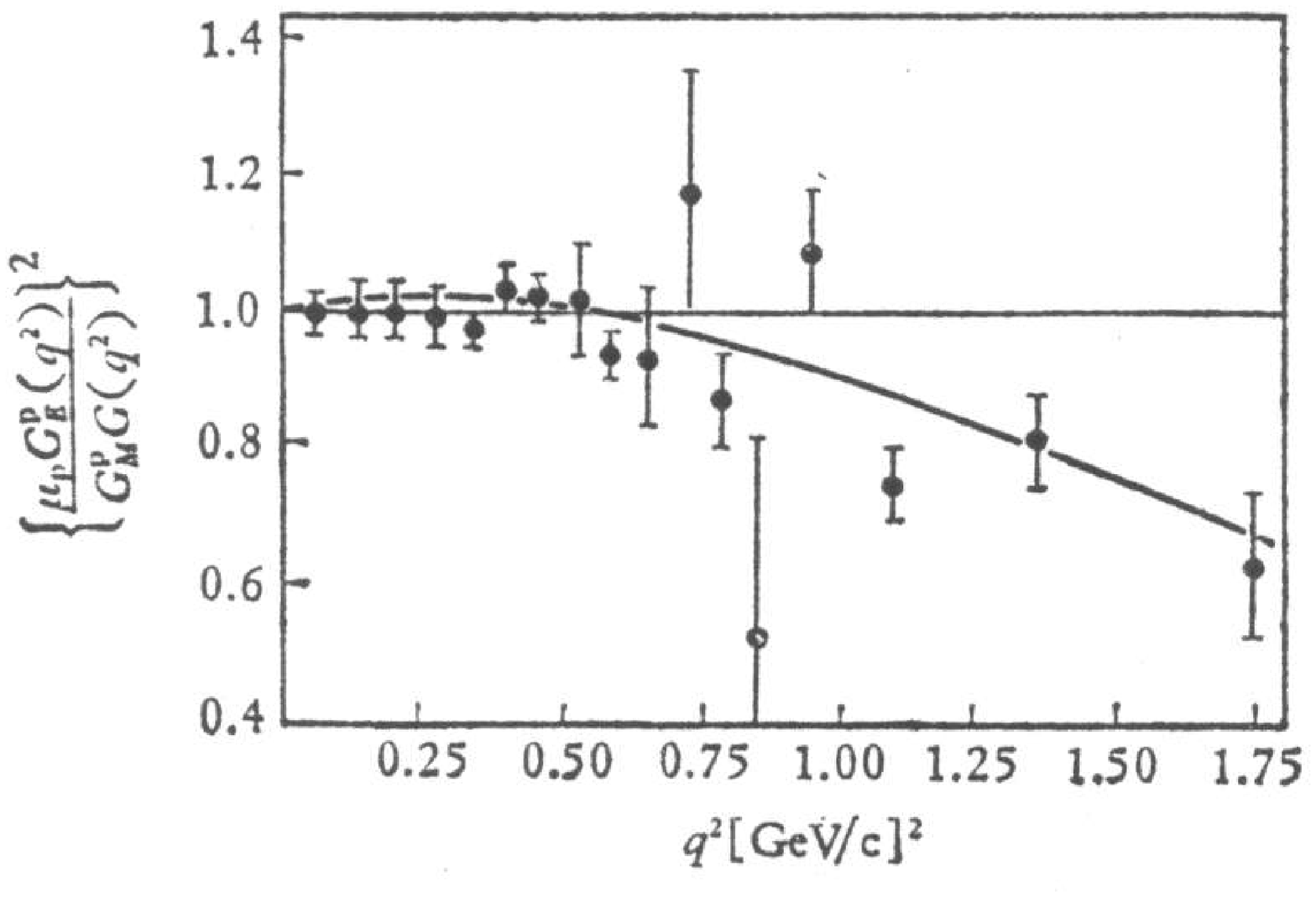}
FIG. 1.
\end{center}
\end{figure}

\begin{figure}
\begin{center}
\includegraphics[width=7in, height=7in]{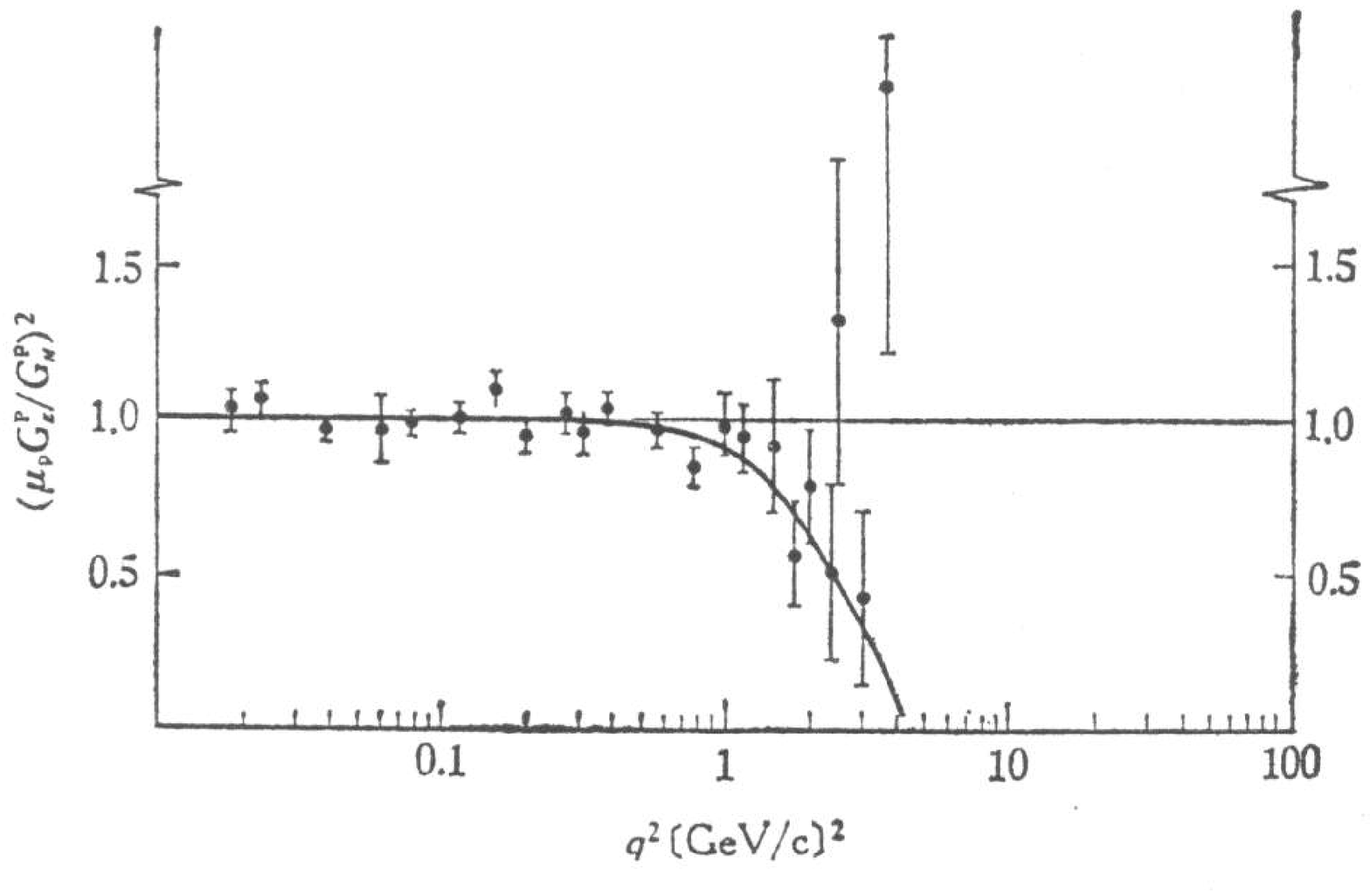}
FIG. 2.
\end{center}
\end{figure}

\begin{figure}
\begin{center}
\includegraphics[width=7in, height=7in]{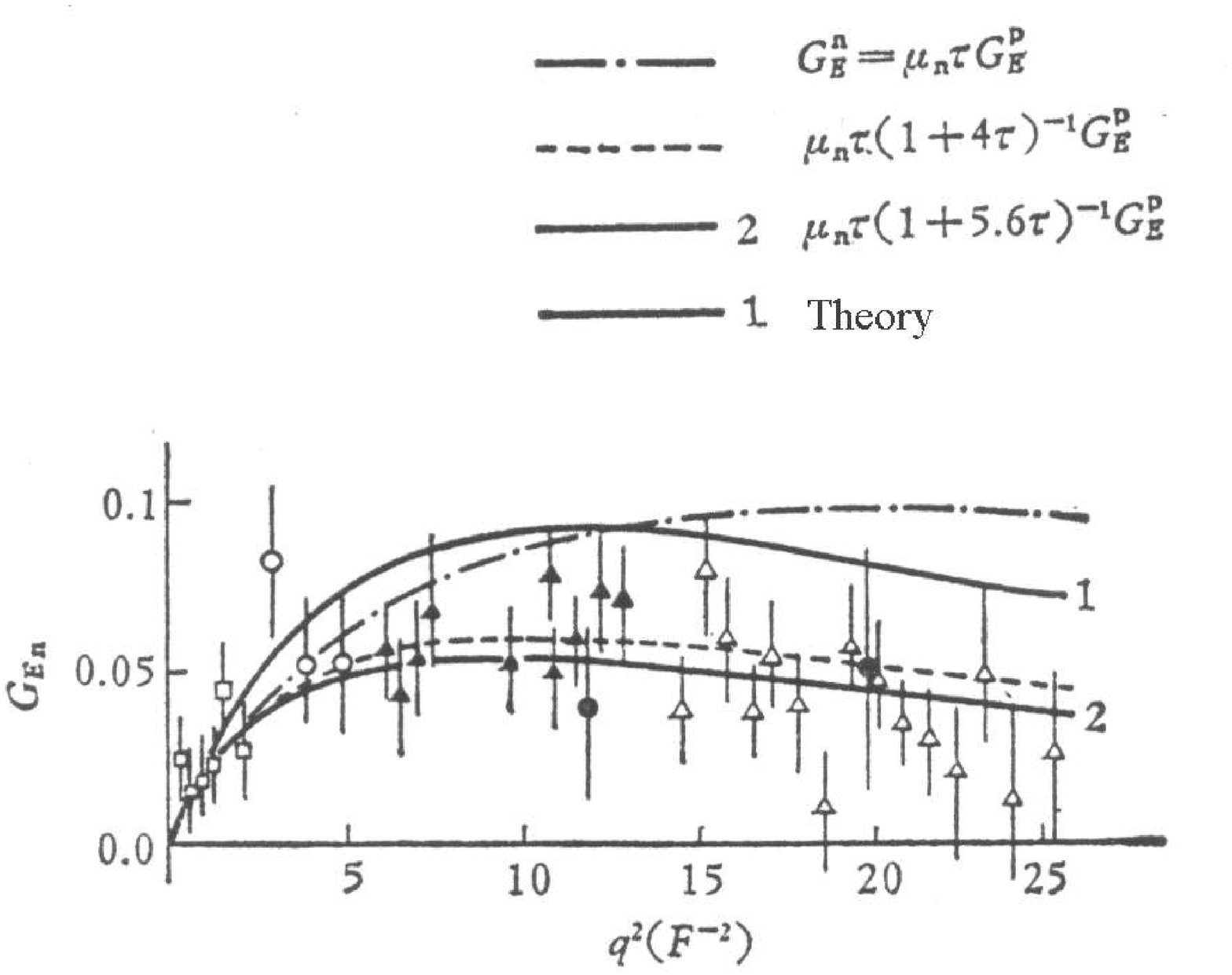}
FIG. 3.
\end{center}
\end{figure}

\begin{figure}
\begin{center}
\includegraphics[width=7in, height=7in]{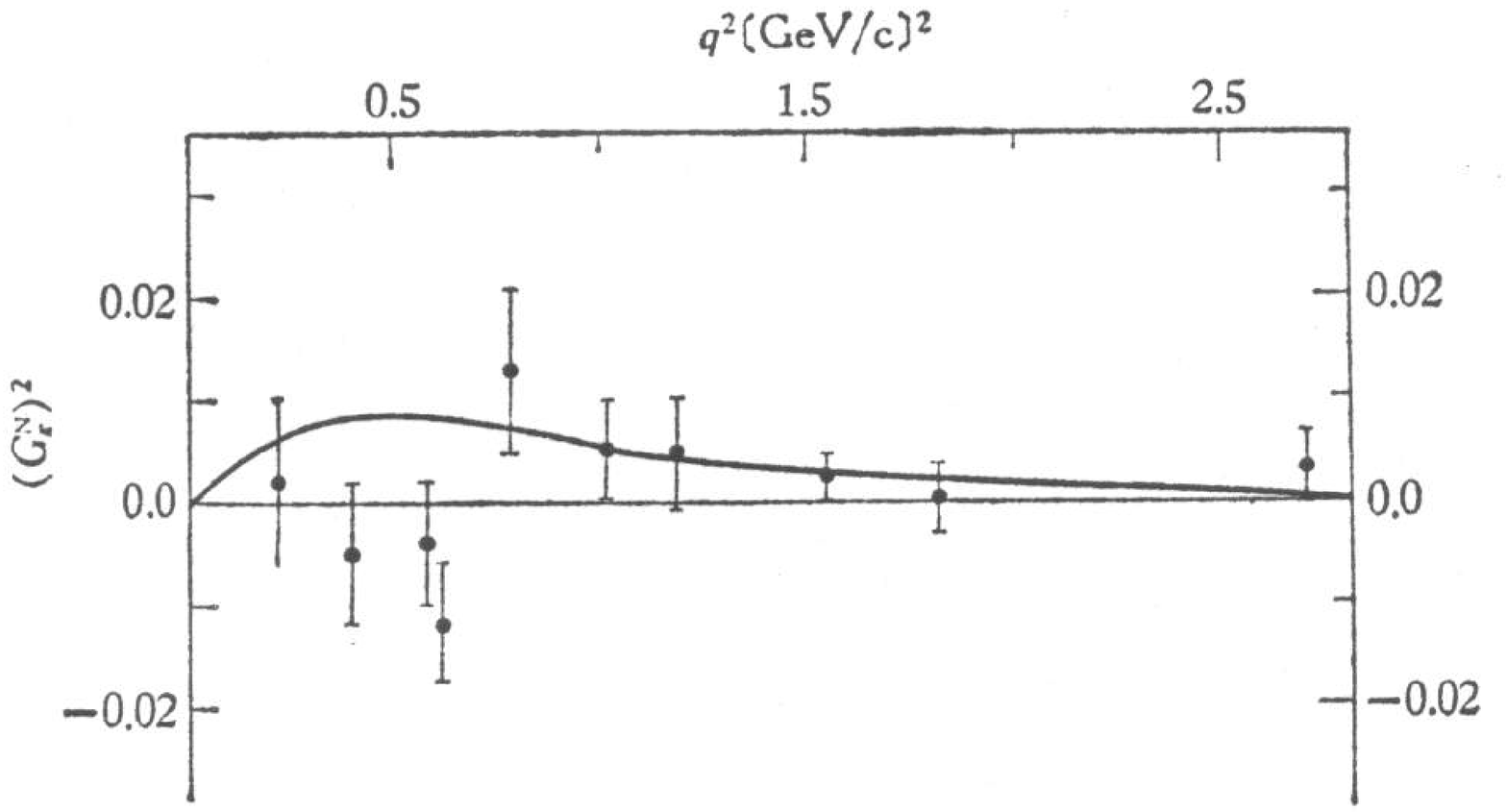}
FIG. 4.
\end{center}
\end{figure}

\begin{figure}
\begin{center}
\includegraphics[width=7in, height=7in]{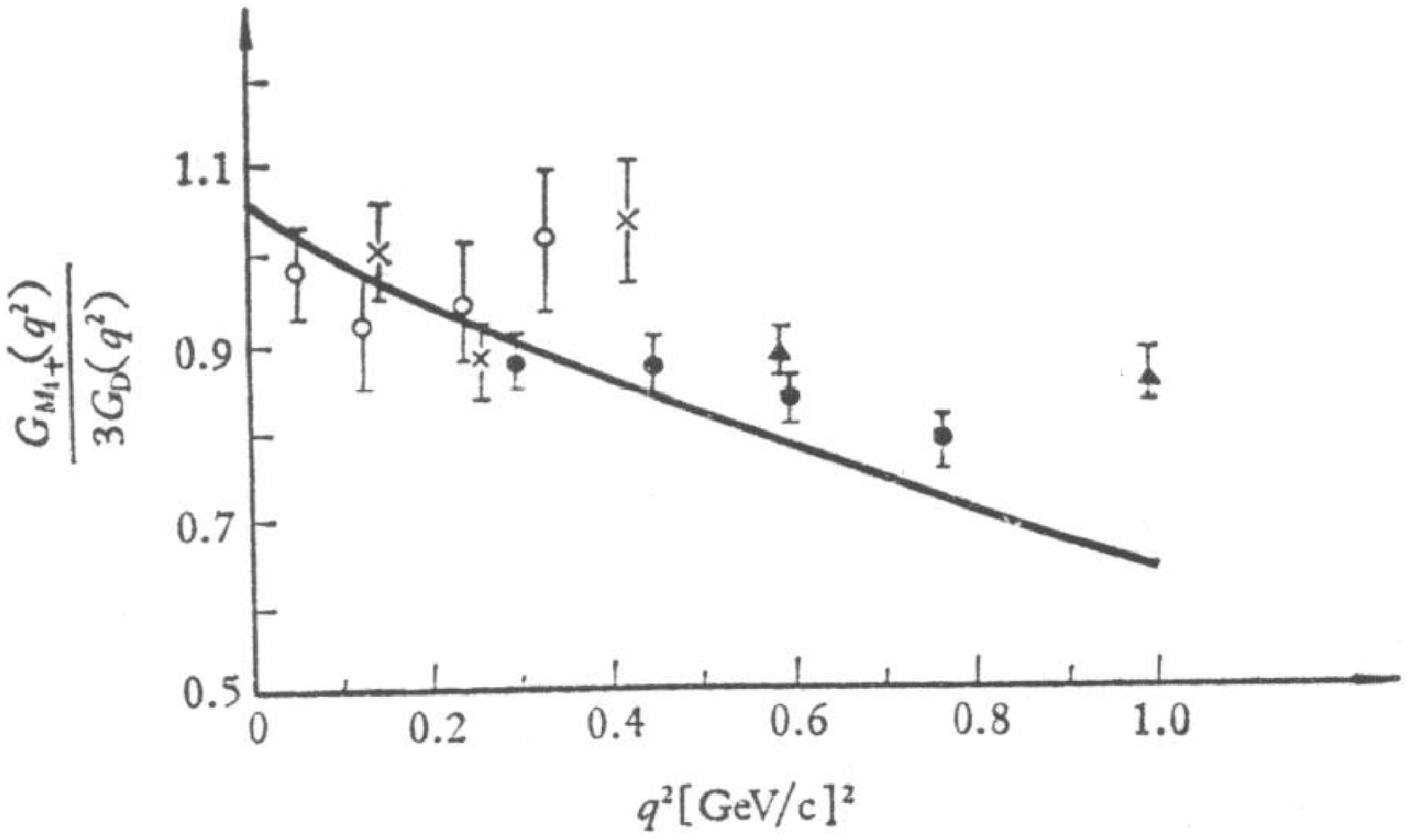}
FIG. 5.
\end{center}
\end{figure}

\begin{figure}
\begin{center}
\includegraphics[width=7in, height=7in]{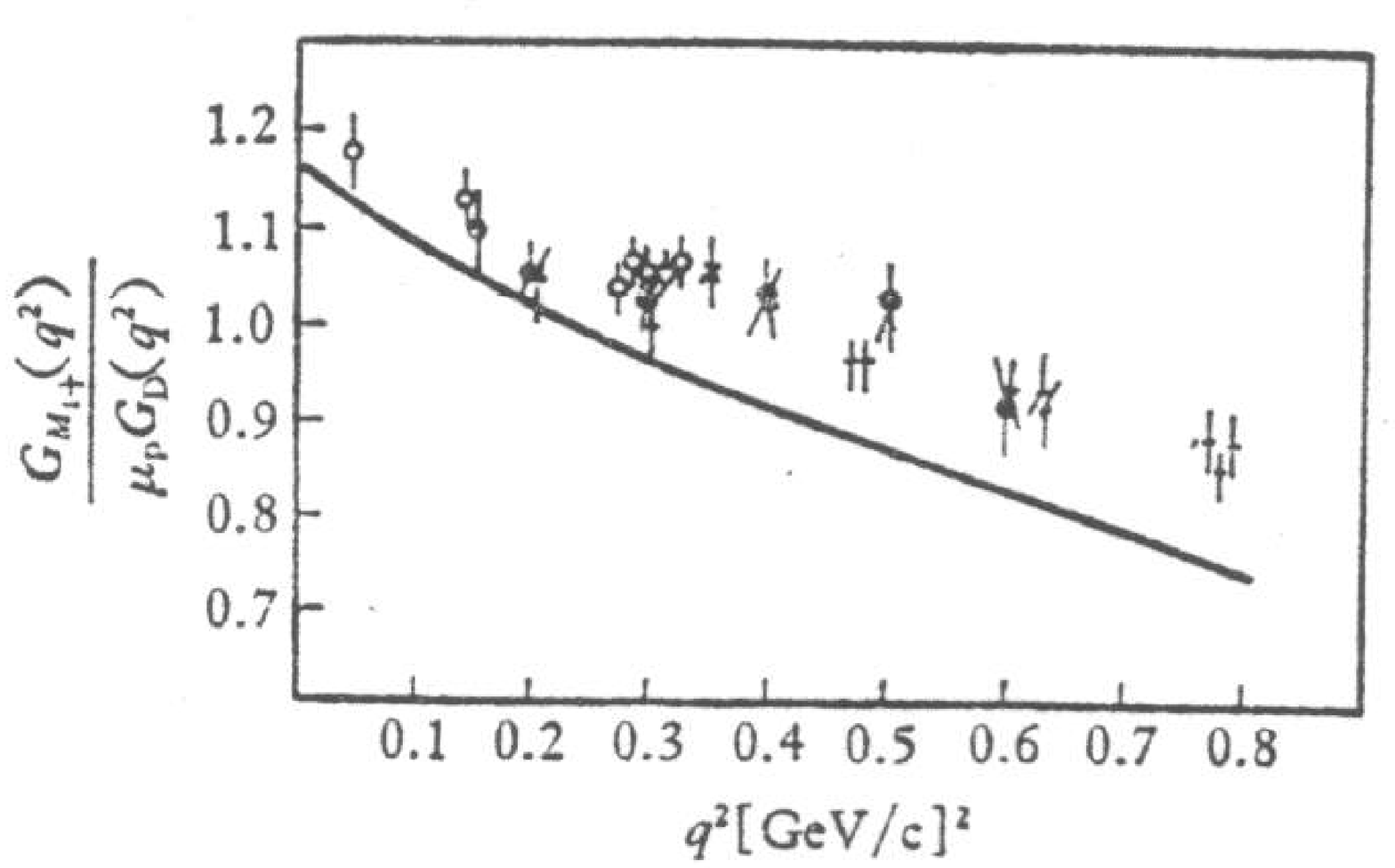}
FIG. 6.
\end{center}
\end{figure}

\begin{figure}
\begin{center}
\includegraphics[width=7in, height=7in]{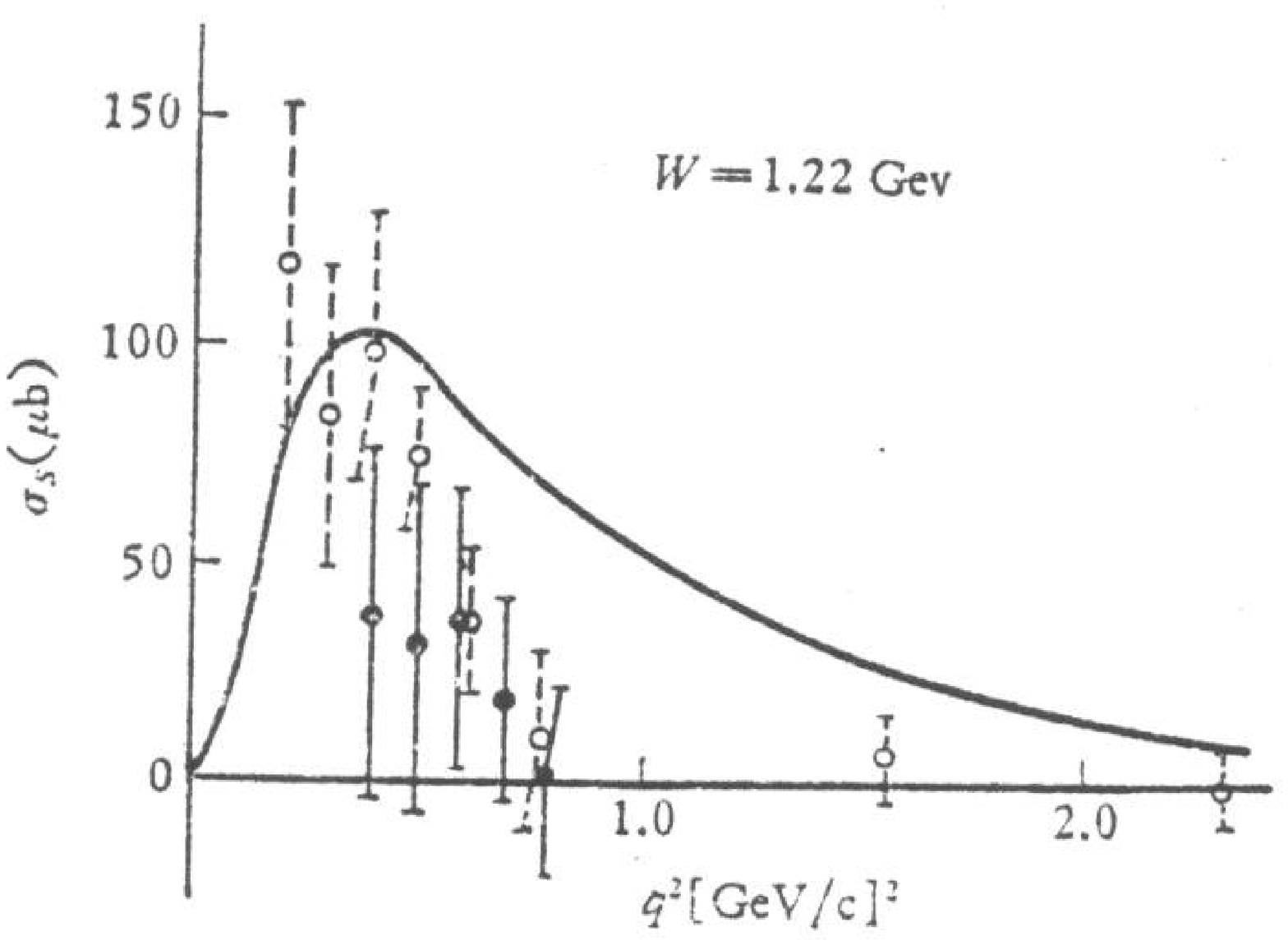}
FIG. 7.
\end{center}
\end{figure}

\end{document}